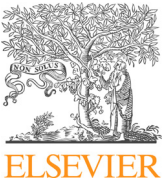
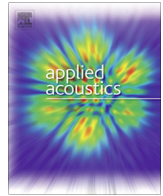

# Two simultaneous talkers distract more than one in simulated multi-talker environments, regardless of overall sound levels typical of open-plan offices

Manuj Yadav\*, Densil Cabrera

*Sydney School of Architecture, Design and Planning, The University of Sydney, Sydney, NSW 2006, Australia*

## ABSTRACT

The irrelevant speech effect (ISE) characterizes detriment to cognitive task performance in the presence of irrelevant speech. This paper examines whether the ISE varies due to the number of simultaneously active nearby talkers (for up to two talkers), or the overall sound level, within the context of a simulated open-plan office. Two experiments were conducted within a climate-controlled chamber that was set-up as a medium-sized open-plan office. The cognitive tasks performed by the participants included the digit recall task, and a writing task, within a room acoustic simulation of realistic multi-talker speech from spatially separated talkers. Within Experiment 1 ($n$ = 60), an increase in the number of talkers from none (T0) to one (T1), and from one to two (T2) simultaneous talkers resulted in statistically significant decline in the digit recall task performances, with effect sizes of 24% (i.e., T1 vs. T0), and 12% (i.e., T2 vs. T1), respectively. The pauses between words during the writing task were similar for T0 and T1, but showed a statistically significant increase within T2 vs. T1, with an effect size of 12%. The findings of Experiment 1 are inconsistent with the maximally distracting status attributed to T1 in some studies, but is consistent with findings in other studies. Within Experiment 2 ($n$ = 62), the cognitive performance in T2 remained largely invariant between 45 and 57 dB (A-weighted sound pressure levels), which represents a typical range of levels within open-plan offices. This finding is somewhat consistent with previous studies, where the ISE has been shown to be invariant over large changes to the overall level of more-or-less spatially-static single-talker speech. However, the current findings provide a more ecologically-valid representation of multi-talker level-invariance of the ISE. In general, these findings have relevance for characterizing auditory distraction within complex multi-talker environments; both in laboratory studies and actual open-plan offices.

Crown Copyright © 2018 Published by Elsevier Ltd. All rights reserved.

## 1. Introduction

### 1.1. Irrelevant speech in open-plan offices

There is sizable evidence now that implicates irrelevant but intelligible speech to many ill-effects in open-plan office sound environments [1]. Despite advances in studying and 'treating' the acoustics of *unoccupied* open-plan offices with the support of a recent international standard (ISO 3382-3 [2]), speech from nearby talkers in adjacent workstations in *occupied* offices is still likely to be intelligible, and hence, disruptive [3]. More generally, nearby talkers in multi-talker environments that are spatially separated represent a source of detriment to performance in certain cognitive tasks [3–6]. The current study considers whether it is the number of talkers, or the overall speech level, that affects task performance, for up to two simultaneous talkers within a medium-sized open-plan office simulation.

### 1.2. The irrelevant speech effect, and the duplex-mechanism of auditory distraction

The irrelevant speech effect (ISE; applicable also to some sounds other than speech) represents a widely accepted framework across many disciplines, wherein auditory distraction is represented as a systematic performance detriment within a goal-driven task, in the presence of to-be-ignored speech that is also task-irrelevant. The main task used to study the ISE has been the serial recall task,

\* Corresponding author at: Room 589, 148, City Road, Sydney School of Architecture, Design and Planning, The University of Sydney, Sydney, NSW 2006, Australia.

*E-mail addresses:* manuj.yadav@sydney.edu.au (M. Yadav), densil.cabrera@sydney.edu.au (D. Cabrera).

which involves recalling a visually-presented, ordered sequence of items (digits, alphabet letters, etc.) from short-term memory [7]. Despite its simplistic, and somewhat rigid nature, the performance in this task is nevertheless linked to many higher-level cognitive functions, such as language processing, problem solving, etc. [8]. Other tasks, such as writing, proofreading, etc., that are more representative of tasks performed in offices have been used in the past within the context of ISE (e.g., [5,9–13], but they are uncommon compared to serial recall tasks, with one difficulty being the generalization of such tasks (e.g., writing, etc.) across different scenarios.

However, the findings from both the serial recall, and tasks for which performance may involve traversing a longer-term semantic network, are nevertheless explicable within the duplex-mechanism account of auditory distraction, which is the leading theory for auditory distraction from a behavioral-cognitive perspective ([7], but see [14] for a summary of the pros-and-cons of the alternative unitary account, and results that challenge the duplex theory). The duplex theory encapsulates two functionally distinct mechanisms – interference-by-process, and the attentional capture [7,15]. As the name suggests, the former is produced when the involuntary processing of to-be-ignored speech competes with the dominant type of processing required in performing the focal task [7]. For instance, while the obligatory yet involuntary processing of order implied in the segmented changing-state speech stream[1] has been shown to disrupt the deliberate order processing required in seriation, with minimal semantic interference (summary in [7,18], although see [19,20]), it is the pre-attentive semantic processing of the irrelevant speech – not its changing-state nature – that activates semantic retrieval processes that compete for similar processes required in semantically intensive tasks (such as writing) to cause performance detriment (reviewed in [21]). With attentional capture, unexpected deviations in the sound stream, such as one's name being called, or other substantial non-acoustic changes (presumably talker loci in multi-talker speech) can also divert attention from the focal task [7]. The magnitude of such deviation effects on the overall ISE has generally been shown to be relatively less than that due to interference-by-process (reviewed in [7]). Furthermore, such attentional deviations have been shown to occur, to varying degrees, regardless of the dominant process involved in the focal task – hence, the functional differentiation of its mechanism from interference-by-process ([7] c.f., [14]).

### 1.3. The level-invariance of the ISE within single-talker scenarios

One of the key characteristics of the ISE in serial recall is that its magnitude has been shown to not vary much over large changes of the overall sound levels – commonly referred to as the *level-invariance* of the ISE in serial recall [22]. To the best of the authors' knowledge, while some studies have investigated the ISE in tasks requiring focal semantic processing, the level-invariance aspect has not been studied directly. Colle [23] showed that ISE in serial recall (of letters) did not vary appreciably over an $L_{A,eq}$ (A-weighted equivalent sound pressure level) range of 40–76 dB, with the impairment due to ISE disappearing only at a very low level of 20 dB, which was 12 dB above the listeners' hearing threshold. Ellermeier and Hellbrück [22] showed level-invariance in the ISE using a serial digit recall task for signal levels of $L_{A,eq}$ 60–75 dB (irrelevant speech and staccato music), mixed with relatively steady-state background-noise ($L_{A,eq}$ ∼41 dB; Experiment 1 in [22]), and further concluded (Experiments 2 and 3 in [22]) that the magnitude of the ISE in serial recall grows from no effect to the maximum, over a speech-to-noise ratio of 16 dB. The speech-to-noise ratios were achieved by using varying levels of pink noise, up to a maximum $L_{A,eq}$ of 77 dB, mixed with a speech $L_{A,eq}$ fixed at 65 dB. Tremblay and Jones [24] showed level-invariance in the ISE of serial recall over a $L_{A,eq}$ range of 55–85 dB, and Schlittmeier et al. showed level-invariance in serial recall and mental arithmetic tasks using two levels of speech (35 dB and 55 dB $L_{A,eq}$) played at either low or high intelligibility to participants [25].

Collectively, the ISE in serial recall has been shown to be level-invariant over a $L_{A,eq}$ range of 40–85 dB at several speech-to-noise ratios in several laboratory-based studies (e.g., [22–25]). However, all of these studies had relatively simplistic methods for presenting the irrelevant sound environments, which included repetitive speech from a single talker with a spatially fixed location, and especially high sound levels (with the exception of [25]). In terms of levels, the ambient A-weighted levels (noise and speech combined) in typical open-plan office environments are recommended to be within a $L_{A,eq}$ range of 40–55 dB according to a recent French standard, depending on the prominent workplace activity [26]. Furthermore, noise levels ∼45 dB $L_{A,eq}$ [27] have been subjectively reported as too loud by office workers, and single-talker speech environments with steady-state background-noise of ∼48 dB $L_{A,eq}$ have been reported as annoying [9]. Overall, it is not immediately clear if the findings of level-invariance in the ISE in serial recall also hold for sound environments with more realistic sound levels, and spatially dynamic, and moreover semantically valid, multi-talker speech that is common to open-plan offices (i.e., ISE in semantic tasks).

### 1.4. The ISE within multi-talker environments with spatial separation of talkers

Jones and Macken, in one of the very few studies of the ISE in serial recall within multi-talker environments, conducted several experiments where the number of simultaneous talkers, and their spatial location, were controlled as independent variables [4]. In Experiment 4 within their study, which is the most relevant here, serial recall performance was significantly reduced when six spatially separated loudspeakers were used to simulate six talkers (i.e., one talker per loudspeaker), compared to when all loudspeakers played the combined speech *babble* from these six talkers. Thus, the ISE in serial recall was shown to increase due to a form of spatial unmasking, or segmentation, where spatial separation of talkers is likely to have increased the possibility of localizing individual voice segments (including their changing-state spectral and temporal cues), and hence, increase the overall changing-state. The authors, however, do not provide any objective data, e.g., in the form of the sound pressure level, etc. of the 'loudness-matched' stimuli that were used in the experiments. Moreover, the speech stimuli used for both the babble, and spatially separated voices played over six loudspeaker conditions, consisted of looped 20 s speech samples [4], which represents limited ecological validity in terms of simulating open-plan office environments.

More recently, Haapakangas et al. [3] studied the ISE using several cognitive tests within a realistic open-plan office simulation, where four individually controlled loudspeakers played non-repetitive speech at levels more representative of open-plan offices (50–53 dB $L_{A,eq}$ at workstations, due to speech from the nearest loudspeaker 2 m away; see Table 2 in [3] for details). Only one loudspeaker (i.e., simulated talker) was active at any time, with the loudspeaker order randomized. In this regard, the practicality of using a single active talker within an already complicated exper-iment design with multiple independent factors (e.g., acoustic absorption, background-noise) partly justifies the authors' choice of not testing more than one simultaneously active talkers. However, as acknowledged within Haapakangas et al. ([3], p.14), the findings of Jones and Macken [4] imply that having more than one simultaneously active talkers that are spatially separated may lead to increased disruption in cognitive performance for multi-talker speech.

---
[1] Note that the concept of 'segments' in changing-state speech shares many similarities with idea of the perceptual organization of sound elements, or *streaming* in auditory scene analysis (ASA) [16]. While some ASA concepts (specifically auditory streaming) are used in the current study, arguably most of the ASA framework has been based on studies involving selective or focused attention to auditory stimuli [17]. In contrast, the hallmark of the ISE is in addressing situations where attention to ambient (irrelevant) sounds is neither desired nor recommended. As such, generally, the ISE framework seems better suited to study some aspects of multi-talker scenarios like those in open-plan offices.

Intuitively, the combined 'babble' of many talkers that is either collocated, or spatially spread, such as in an office environment, is not likely to be disruptive, and may in fact provide beneficial sound masking [28]. Nevertheless, highly intelligible speech from nearby talkers is still likely to be disruptive [3], against the background of steady-state noise, and/or more babble-like speech from relatively distant talkers. The characteristics (e.g., spectrum, number of talkers and spatial distribution, etc.) of this pre-babble, and disruptive multi-talker speech from nearby talkers are still relatively unexplored, at least with respect to the ISE. Furthermore, due to other psychoacoustic benefits from spatial separation (summary in [29]), there may be separate segmented streams from nearby talkers in the pre-babble multi-talker speech: a scenario shown to be more disruptive than a single segmented stream [8].

Yadav et al. [6] recently explored one such characteristic of the pre-babble multi-talker speech, in terms of the number of simultaneous talkers within close range that contribute to task disruption. This study was similar in design to the study by Haapakangas et al. [3], at least in terms of using individually controlled loudspeakers as speech sources with non-repetitive and quotidian speech, and approximating the looks and climate of a medium-sized open-plan office. The results showed that compared to single-talker speech, as used in Haapakangas et al. [3] and silence, multi-talker speech with two and four simultaneously active talkers led to performance decline in some of the cognitive tasks, and was also perceived to be subjectively more distracting by the participants, regardless of whether there was additional broadband sound masking [6]. The decline in cognitive performance shown with multi-talker speech, with more than one spatially-separated talkers, was similar to the findings of Jones and Macken [4], although with a more ecologically valid experimental scenario, while the increase in subjective auditory distraction with the talker numbers was a new finding [6].

However, there were some limitations in terms of contextualizing the findings of the study by Yadav et al. [6], within bulk of the previous literature of the ISE. In particular, the cognitive tasks that were used did not include the more traditional serial recall task, which limits comparisons with previous studies – a hindrance to arrive at robust findings based on combining results from multiple studies, as highlighted in a recent review [30]. Instead, the digit span task, which has several methodological differences from the more traditional digit recall task, was used [6]. For the digit span task, while there was a trend showing performance decline with increasing number of talkers, statistical significance was not reached. This limitation in terms of using non-traditional cognitive tasks within the scope of the ISE is addressed in the current study by using the digit recall task, with a larger sample size. Moreover, the findings relating to multi-talker speech in Yadav et al. [6] were confounded by the overall sound levels (amongst other factors) also increasing with increasing number of talkers. Hence, it was not clear whether the findings in [6] were attributable simply to the increase in the overall sound levels, or to the number of simultaneously active talkers; where the latter may depend on several other factors such as speech content, spatial location, etc. Such considerations are more directly addressed in the current paper, the aims of which are detailed in Section 1.5.

### 1.5. Aims of the current study

In the current study, the effect of the number of simultaneous talkers up to a maximum of two, and the overall sound levels on the ISE is studied over two experiments. The overall aim is to both replicate, and extend previous findings [3,4,6,22], within the context of multi-talker speech environments representative of a typical medium-sized open-plan office that has spatially varying talker number and locations, realistic sound levels, and non-repetitive, semantically rich speech. The tasks used included the more common digit recall task to enable a more direct, and comprehensive comparison with previous studies, which was not possible with a previous study [6]; and a writing test that involves a semantic aspect in its performance, while also more closely representing typical office tasks.

## 2. Methods

### 2.1. Room acoustic simulation

Instead of using a loudspeaker-based system, as done previously in similar research, a headphone-based system is used here for playing multi-talker speech environments within the simulated room acoustics of an actual room. The main reason was to enable testing multiple participants simultaneously for the same multi-talker acoustic environment, which would otherwise vary with the location of the loudspeakers.

A CAD model was created of a 60.6 m² room (8.9 m × 6.9 m × 2.6 m; mid-frequency $T_{30}$ = 0.3 s) that was used in previous research [6]. The actual room visually resembles a medium-sized open-plan office, and the CAD model included all the furnishings of the room, including desks, chair, partitions (1.5 m × 1.5 m each), cabinets, computer screens, etc. This model was imported into the ODEON software [31], wherein its acoustic properties were closely matched with those of the actual room. The criteria here were to match (within 5% of the respective one just-noticeable difference values [32]) the reverberation times ($T_{30}$) and clarity indices ($C_{50}$) of the room impulse responses (RIRs) measured at several locations in the actual room, and those simulated for the same locations within ODEON for the modelled room.

Room acoustic simulations were then run within ODEON to derive the RIRs for a binaural receiver at the workstation location R, as seen in Fig. 1, with four sources representing talkers at adjacent workstations locations, all at heights of 1.2 m; similar to previous research where loudspeakers were used to simulate talkers [6]. The binaural receiver was modelled in ODEON as a person sitting on a chair, and the sources were assigned voice directivity characteristics (BB93_Normal_Natural.So8 in ODEON) to simulate a person speaking with a normal vocal effort (as defined in Table 4 of ANSI 3.5-1997 (R2007) [33]).

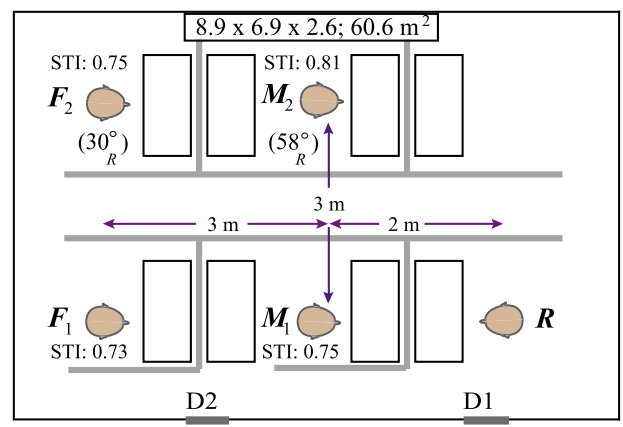

**Fig. 1.** General layout of the room that was simulated, showing the angular separation of the receiver (R) from the talkers ($F_{1\&2}$, $M_{1\&2}$) along the length, and speech transmission indices (STI; calculated in ODEON) and the distances per talker-receiver configuration. Doors are indicated (D1&2), and the thick grey lines show the office partitions (1.5 m × 1.5 m).

### 2.2. Stimuli

The speech material used consisted of edited portions from scripted dialogues, which were recorded (44.1 kHz sampling rate) in an anechoic room by 12 trained actors, over several sessions. Each session included 2 seated actors separated by 2 m, each wearing a DPA d:fine 4066 headset microphone (Allerød, Denmark) that was positioned 7 cm from the centre of lips, to the right side of their faces. The actors were instructed to ensure that the spoken text resembled a real conversation, and not a script being read, with rigorous quality control maintained throughout. The scripts chosen had a talk-show format based on diverse current affairs topics. Each actor was assigned almost half of the speaking, as both the interviewer and the interviewee at set points in the script.

In the following, the signal processing was done in MATLAB, unless mentioned otherwise. Voiced segments from the recordings were separated algorithmically (2–50 s long), and their octave-band spectra converted, using a finite impulse response filter, to match the speech spectra representing 'normal' vocal effort, using values specified in Table 1 of ISO 3382-3 [2] (values for directional source; based on [33]). These spectrum-adjusted segments are referred to as $seg_{ANSI}$ in the following. To create the sound environments used in the experiments (except silence, referred to as **T0** in the following), contiguous $seg_{ANSI}$ from four talkers (two per gender) were chosen, where none of the talkers were in a dialogue with the others. The $seg_{ANSI}$ were first arranged into 4-channel (one per talker) files (Fig. 2), where there was either one active talker (**T1**) or two simultaneously active talkers (**T2**). The order of active talkers in the 4-channel files was algorithmically randomized, and a variable amount of silence (randomized to be between 0 and 4 s) was added between the $seg_{ANSI}$. This additional silence was distinct from the natural pauses that may occur while speaking, which were preserved.

The resulting 4-channel files were imported into ODEON, calibrated, convolved with corresponding RIRs (for receiver *R* in Fig. 1), and converted into a binaural format using head-related transfer functions from Subject_021 of the CIPIC database [34]. A headphone correction filter was used within ODEON for the headphones used in the experiments (Sennheiser HD 600, Wedemark, Germany). Separate binaural files (12–15 min each) were created for the experimental conditions (Section 2.3), where there was no repetition of speech content. Furthermore, the four separate speech streams used here resembled separate telephone conversations, where only one side of the conversation is heard by the listener – a scenario that has been reported by office occupants as one of the most distracting [35].

### 2.3. Experimental conditions

Two separate experiments were run with different combinations of the experimental conditions i.e., T0, T1, and T2. The headphone reproduction levels for the experimental conditions were based on $L_{A,eq}$ values of the sound stimuli recorded for a 15-minute duration, using the ears of the Neumann KU100 dummy head (Berlin, Germany) wearing the headphones. For T0, the ambient A-weighted level in the experiment room was approximately 31 dB (with and without headphones, also see Section 2.4). The earcups of the Sennheiser HD 600 units have an open-back circumaural design, but there was no appreciable change in the recorded levels for T1 and T2 due to the ambient sound leaking in, when tested with the typical use of the keyboard and mouse for the experimental tasks. The STI values for each talker-receiver configuration are given in Fig. 1. Note that for the experiments, the STI values are indicative only, as multi-talker STI is currently not possible [6,36].

In **Experiment 1**, the experimental conditions included silence (T0), and speech with one (T1), or two simultaneous talkers (T2; Fig. 2). For T1 and T2, the overall A-weighted levels over the headphones were set at 49 dB, and 51 dB, respectively. These were representative of typical levels (or, targets in terms of design) in occupied offices that involve mostly collaborative work but also involve the need for individuals to concentrate, call-centres, etc. [26]. Furthermore, the speech-to-noise ratio for T1 and T2 is 18 dB, and 20 dB, respectively, where the corresponding value for T1 within comparable experimental conditions was between 15 and 18 dB in Haapakangas et al. [3]. Limiting the talker numbers to two simultaneous talkers represents a reasonable, and logistically viable starting-point for an ecologically-valid scenario in medium-sized open-plan offices, and perhaps even in larger offices, when the nearby talkers are generally intelligible in favorable scenarios, e.g., when there is limited occupancy, and/or the ambience beyond the nearby workstations is appreciably quieter. It is expected that the results can better inform future research on other psychoacoustical aspects of multi-talker speech environments (such as babble speech [4]), potentially with more talkers.

For **Experiment 2**, the experimental conditions included T2 played over a range of gain values (+6 dB, +3 dB, −3 dB, −6 dB), relative to A-weighted level of 51 dB used in Experiment 1 for T2, i.e., a range of 12 dB from 45 to 57 dB, which represents typical (or, target) levels in almost all types of open-plan offices (based on [26]).

### 2.4. Procedure

**Experiment 1** had 60 participants (31 F; age range: 18–55, mean: 25.1, standard deviation: 6; 21 native Australian English speakers), and **Experiment 2** had 62 participants (31 F; age range: 18–55, mean: 25.6, standard deviation: 7; 30 native Australian English speakers). None of the participants reported hearing impairment. The duration of each experimental session was two hours, which included time for instructions, practice, experimental tasks, breaks, and debriefing.

The experiments were conducted in the same open-plan office as the acoustically modelled room (Fig. 1). The climate in the room was maintained using underfloor air distribution (22.3 ± 0.3 °C), which had an operational A-weighted noise level of ∼31 dB (i.e., ambient level for T0). The luminance at the workstations was maintained at 500 lx, measured near the active iMac screens used for the experiment. The ceiling of this room has acoustic tiles, floor is carpeted, walls have acoustic diffuser panels, with plants and some office paraphernalia present throughout; lending a heightened sense of working in an actual office.

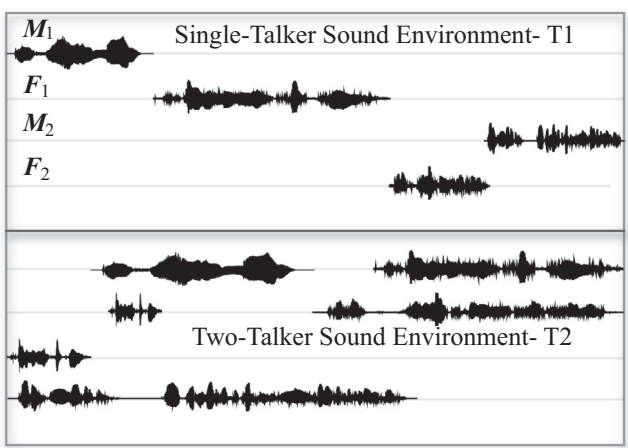

**Fig. 2.** An example of 4-channel speech for T1 and T2 that was binaurally auralized for *R*. The talkers ($F_{1\&2}$, $M_{1\&2}$) are labelled in accordance with the layout in Fig. 1.

Most experimental sessions had three participants, who were allocated workstations farthest from each other, and separated by multiple partitions. Each workstation had an iMac computer, and very quiet keyboards and mice (on thick cushioned pads) were provided to minimize noise due to their usage. The experimenters were located in an adjoining room after setting up the participants, which simply included showing the participants how to correctly wear the headphones (worn throughout the session). The rest of the experiment was self-sustained, and began with a 15-minute instructional video that gave a comprehensive overview of the tasks to be performed, their order, information regarding not disturbing the other participants, contacting the experimenter, etc. Additionally, detailed instructions were provided as on-screen text before every experimental task. Participants also filled in a noise sensitivity questionnaire before the experiment started, which had items on a continuous scale (Completely Disagree – Completely Agree) [37].

The experiments included the serial digit recall task, and a writing task, both run within a graphical user interface (GUI), programmed in the software Max (Cycling '74, Walnut, USA), which was also used for data management and audio signal processing. While it was mentioned that there may be certain sound environments played over the headphones during the tasks, no instruction was given to attend to those sounds.

For the *digit recall task*, eight digits from 1–9 (randomly sampled, with no repeats), were visually presented in a box (72px font) in the centre of the GUI. Digits were presented for one second each, with an inter-digit duration of one second. After all the digits were presented, the participants recalled the digits (no time limit), in order, on a 4 × 3 numeric grid (last row had only 0 in the middle) using mouse clicks, inputting 0 for every digit that was not recalled. The selected digits disappeared from the GUI (except 0), and after selecting eight digits, the participants pressed a button to continue further. The sequence of presentation and recall of the eight digits was referred to as a *trial*, and twelve such trials constituted a *block*, where each block tested a certain experimental condition (i.e., T0, T1, or T2). The participants first did a practice run of the digit recall task (12 trials, or 1 block) in silence, followed by the experimental blocks (3 in Experiment 1, 4 in Experiment 2), without any break in between. The order of the sound environments was randomized across the participants, and the sounds played continuously in each experimental condition (except, of course, in T0).

The *writing task* followed the digit recall task, beginning with a practice session. The GUI showed a figure on the left side of the screen, and a blank writing space on the right. There was a backwards time-counter, and a user-interface button to control the flow of the experiment. The instructions were to interpret the information in the figure, and to write a professional report with as many words as possible without errors (no autocorrect facilities). The duration for each trial (different sound environment per trial) was set at 10 min (3 in Experiment 1, 4 in Experiment 2). The participants had the option of finishing a trial before the allotted 10 min, although none of the participants used this option; at the end of 10 min, keyboard input stopped. The information presented in the figures changed per trial, and consisted of bar– and pie-charts, tables, diagrams of processes (e.g., rain formation cycle, etc.), and a combination thereof, along with some text that contextualized the figure. The figure contents were from online samples of academic report writing for an English proficiency test (IELTS; https://www.ielts-exam.net/). The difficulty of each figure was kept approximately similar. The figures chosen and the order of the experimental conditions were randomized per trial.

### 2.5. Data analysis

All data processing and statistical analyses were done using the software R. For the *digit recall* task, only 10 trials per block were analyzed, after removing the first and last trial. This was done to minimize any residual effect on task performance due to the sound environments changing for each block. A correct response per digit per trial was registered for the digits recalled in the exact serial order of their presentation, which was then averaged over all the trials (referred to as $digRecall_{Error}$), and used as the response variable in statistical analyses. The relative change in error rates [17] was used to indicate the effect size, calculated as $[(E_{soundEnv1} - E_{soundEnv2})/E_{soundEnv2}]$, where $E_{soundEnv[1,2]}$ represents the digit recall errors in the respective sound environment. For instance, $[(E_{T1} - E_{T0})/E_{T0}]$ represents the effect size of the relative change in error rates between T1 (one-talker scenario) and T0 (silence).

For the *writing* task, several performance metrics (response variables) were derived from the participants' written text. These included the number of characters written, deleted, pauses greater than 2 s duration, spelling errors, and total words written; no qualitative analysis of the text was done. To control for the differences in the total material produced across participants, the response variables for deletes and pauses were adjusted in the following manner. For each writing trial per participant, the total number of deletes was divided by the total characters written, which formed the response variable ($deletes_{Adj}$). Similarly, for each writing trial per participant, the number of pauses was divided by the sum of total characters written, including characters deleted, which formed the response variable ($pause_{Adj}$).

The effect sizes for the performance metrics in the writing task were calculated in a manner similar to those for the digit recall errors. For each performance metric (e.g., characters written, deleted, etc.), the relative change in scores between the sound environments constituted the effect size, which was calculated as $[(P_{soundEnv1} - P_{soundEnv2})/P_{soundEnv2}]$. Here, $P_{soundEnv[1,2]}$ represents the score for a particular metric in the respective sound environment. For instance, for the total characters written, $[(char_{T1} - char_{T0})/char_{T0}]$ represents the effect size of the relative change between T1 and T0.

For both the tasks, mixed-effects models were fit using the function *lme* from the nlme [38] package with the *sound environment* (levels T0, T1, etc.) as the fixed-effect. The random-effect, due to each participant performing the tasks within every experimental condition (i.e., sound environments) – a repeated-measures design– was modelled by allowing both the intercept and slopes to vary for each participant. The goodness-of-fit comparisons of models varying in their fixed-effects were done using chi-square tests on their respective log-likelihood values (note that mixed-models provided better fit than corresponding GLM models). Model parameters of the mixed-effects models were estimated using the restricted maximum-likelihood method. Residuals of the final models met parametric assumptions (i.e., linearity, normality, and homoscedasticity) across all factor levels. Comparisons of the factor levels of the fixed effects was done using orthogonal contrasts (where possible), or *post-hoc* comparisons using the *pmmeans* function from the *lsmeans* [39] package, with Benjamini-Hochberg method corrected *p*-values. Note that factorial designs, with additional between-subject effects of participants' *native language* (native Australian English speakers, or not) and *noise sensitivity* (yes or no, based on a median split on the scores) were also attempted, but not reported here because the models based on such factorial design combinations were not significant (based on log-likelihood model comparison method described above) compared to just the repeated-measures design.

## 3. Results and discussion

For brevity, only the results for the final models (with the best fit) are presented and discussed.

### 3.1. Experiment 1

The total errors in the digit recall task ($digRecall_{Error}$) increased with the number of active talkers in the sound environments, as seen in Fig. 3. This was confirmed in the modelling, where the experimental condition was a significant predictor of $digRecall_{Error}$ compared to the grand mean (i.e., model with just the intercept). Orthogonal contrasts showed that more errors were made in speech compared with silence, i.e., T1 and T2 together vs. T0 ($b = 0.5$, $t(118) = 6.9$, $p < 0.0001$), and more errors were made in T2 vs. T1 ($b = 0.4$, $t(118) = 4.1$, $p < 0.0001$). The latter validates, and further extends the results in Yadav et al. [6], where decline in cognitive performance was noticed with more than one simultaneous talkers within experiments conducted in the actual room that is simulated here. The trend shown in the ISE (for serial recall) here is also consistent with the findings of Jones and Macken, where the performance decline was expected to plateau after a certain number of talkers greater than two [4] (a babble-like state); although more studies would be needed to confirm this within a multi-talker scenario with spatially-separated talkers. In this regard, four simultaneous talkers still induced performance decline in previous studies [6,29].

Due to the methodological differences between the digit span used in Yadav et al. [6], and the digit serial recall tasks used here, a direct comparison of results is not possible. However, since the digit recall task is the most common task for showing the ISE, it is expected that the current results will provide a more useful context for future research, especially in terms of effect sizes, and of course, for comparison with extant ISE in serial recall literature. Regarding the latter, the study by Haapakangas et al., which was similar to the current in terms of the design of the sound environments, compared digit recall task performance in T0 to T1 (Quiet vs Abs_noMask in their study), which was presented as percent correct responses (Figure 6 in Haapakangas et al. [3]). When converted to the error measure used in the present study (Section 2.5), the errors in T1 relative to T0 increased by 24.5% here, compared to 31.5% in their study [3]. These errors (i.e., for T1 vs. T0) are reasonably similar, given the methodological and room acoustics differences in the respective studies. However, serial recall errors in T2 relative to T1 increased by 12% in the current study, which suggests that the decline in cognitive performance may continue, and be exacerbated with two simultaneous talkers compared to one (i.e., T2 vs. T1).

The possible role of spatial separation of talkers in the current results for the ISE in serial recall can be considered a bit further. To begin with, compared to single-talker speech, two spatially-separated (i.e., unmasked) and simultaneously active talkers constitute more speech segments (single-talker to receiver STIs were fairly high: Fig. 1) over a perceptually salient time window. This implies increased segmentation, and spectro-temporal changing-state, and consequently increased distraction. Besides, in T2, the changing-state may extend to factors other than the spectro-temporal segmentation that generally characterizes speech [4]. These include increased segmentation for T2 vs. T1, in terms of the dynamically changing talker locations over time (i.e., spatial segmentation), and to an extent the gender of the active talkers (with additional spectral changes), etc. For example, consider $F_1$ $M_1$ $F_1$ $M_2$ $M_1$ $F_1$ $F_2$ $M_2$ and $(F_1M_1)$ $(M_1M_2)$ $(M_1F_1)$ $(F_2M_2)$ as the talker orders for T1 and T2, respectively. Here, for every change in the talker order in T1 (e.g., $F_1$ to $M_1$), there are two changes in the order in T2 (e.g., $(F_1M_1)$ to $(M_1M_2)$), leading to an increased degree of the changing-state and the associated performance decline. The discussion here can also be buttressed using an auditory scene analysis [16] perspective. Briefly, increased uncertainty in the talker location, gender, sentence lengths, and moreover, increased energetically unmasked speech implies more (and more frequently updating) auditory streams in parallel with T2 than T1; more auditory streams have been shown to lead to increased errors in serial recall for both relevant and irrelevant speech (reviewed elsewhere [8]).

For the writing tasks, the pauses greater than 2 s ($pause_{Adj}$; see Section 2.5) did not increase much from silence to the single-talker scenario, but increased from the single-talker to the two-talker scenario by 12.3% (Fig. 4). Here, the effect sizes are more easily interpretable than the values on the ordinate. None of the models with other measures of the writing task reached significance. The experimental condition was a significant predictor of $pause_{Adj}$, compared to a model with just the grand mean ($\chi^2 (2) = 7.4$, $p = 0.02$). Pairwise comparisons confirmed the trend in Fig. 4, where a non-significant difference between T0 and T1 ($p = 0.9$) but significant differences between T0 and T2, and T1 and T2 (both $p = 0.04$) were found. This shows that the participants paused more times in two-talker speech than in single-talker speech or silence. While this is a new finding in the context of semantic interference by multi-talker speech, there are a few points to consider.

Since the involuntary processing of irrelevant speech is likely to interfere with the several semantic focal processing aspects involved in writing [12,21], it is somewhat surprising to not see a significant effect from T0 and T1, with a significant effect between T1 and T2. The most plausible explanation of these results is that the more frequent, largely unpredictable spatial shifts in talker locations over time, and the concomitant gender changes in the active voices that are highly intelligible, generated more instances of attentional capture in T2 relative to T1 – hence, more pauses in T2 vs T1. However, the validity of such an argument needs future studies, which should also include testing the variation in other relevant response variables for writing tasks (e.g., number of characters written, etc.), for which process-based disruptions have been noticed in previous studies where similar (though not the same) writing tests were used, albeit without spatially-separated multi-talker speech [10,12]. Overall, pending future studies, it seems more likely that the attentional capture mechanisms have more to do with the current finding of increased pauses between T2 and T1. Moreover, the role of the number of voices and their spatial location in terms of semantic interference cannot be entirely ruled out in multi-talker speech in office scenarios, or in experiments with a variation on the writing tasks, and longer durations, than used here; more studies are needed to study the role of both interference-by-process, and attentional capture mechanisms in realistic writing tasks.

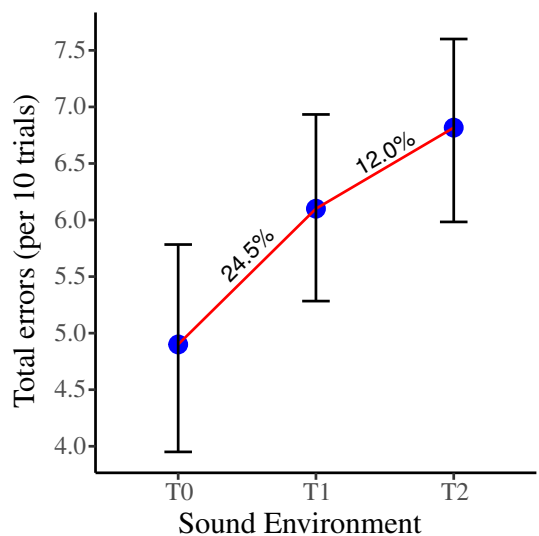

**Fig. 3.** Experiment 1 results for the digit recall task. T0 represents no talkers, T1: one active talker at a time, and T2: two simultaneously active talkers. Mean values shown, along with the standard error (calculated using a non-parametric bootstrap method). The percentage values represent the effect size per comparison of sound environments.

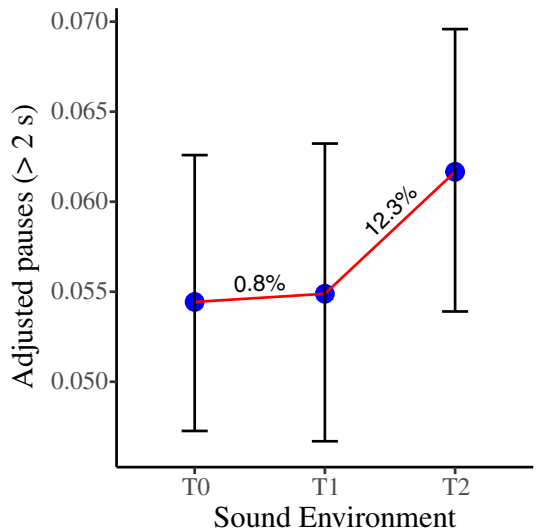

**Fig. 4.** Experiment 1 result for the adjusted pauses (see Section 2.5 for the calculation) greater than 2 s in the writing task. T0 represents no talkers, T1: one active talker at a time, and T2: two simultaneously active talkers. Mean values shown, along with the standard error (calculated using a non-parametric bootstrap method). The percentage values represent the effect size per comparison of sound environments.

There are several methodological limitations in the writing task used here, which include the relatively short duration and the somewhat rigid nature of the writing task, the possibility of some variation in the difficulty levels in the figures used to generate the writing (despite the best efforts by the experimenters to avoid this), etc. While it is not possible with the current methods to determine the effect of such limitations on the current results, the experimental nature and the associated scope of these results must be acknowledged when interpreting them within the context of writing in actual open-plan offices. Furthermore, it must be noted that most of these limitations are more-or-less implicit in selecting a writing task that is suitable within experimental scenarios (see, for instance, writing tasks used in [10–13]). Hence, characterizing the various dynamic and complex aspects of writing [13] in actual offices will benefit from future studies that more directly explore the interaction of some of these aspects, while improving upon the ecological-validity for the tasks performed and the sound environments simulated.

To summarize, the current results can be explained by extending the ISE framework within the context of multi-talker speech, in a manner more-or-less consistent with expectations from psychoacoustic literature (reviewed in [6]) and the cognitive psychology literature [7,21]. It involves considering the effects of spatial shifts in the loci of speech from many simultaneous talkers, where the individual talker's speech has spectro-temporal segmentation, and the speech from many talkers combined is further segmented due to spatial separation and the individual traits of the talkers (i.e., gender, etc.) – *spectro-temporal-spatial* changing-state [4] to cause performance decline in serial recall. Furthermore, the unexpected changes in spatially-dynamic multi-talker speech increases in its capacity to capture attention away for semantically intensive tasks more regularly (e.g., increased number of pauses while writing). Within the extended ISE framework, increased segmentation and attentional capture with (a) two simultaneously active talkers, and (b) a shifting order of talkers, posit more distraction than a single talker. Adding more talkers may eventually lead to spectro-temporal-spatial babble. More research is needed to study the role of semantic (or other) interference mechanism in multi-talker compared to single-talker scenarios, which is not possible given the methodological limitations in the current experimental design, especially the writing task.

### 3.2. Experiment 2

In this experiment, none of the measures in any of the tasks reached statistical significance. For instance, for the digit recall task, modelling the total errors with the sound environment as a predictor did not significantly improve the fit, compared to the grand mean ($\chi^2$ (4) = 6.4, $p$ = 0.2). Hence, for a gain change of 12 dB in the T2 sound environment (−6 dB to +6 dB in 3 dB steps), relative to a sound pressure level of 51 dB (i.e., a 45–57 dB range) used in Experiment 1 for T2, the ISE remained invariant. The mean $digRecall_{Error}$ for the gain values of −6 dB, −3 dB, +3 dB, and +6 dB, relative to 51 dB used in Experiment 1 for T2, were 7.2, 6.9, 6.7, 6.9, respectively. With the performance in all the T2 sound environments combined, $digRecall_{Error}$ was 6.9 ± 0.2, which is very close to the performance in T2 in Experiment 1 (6.8 ± 0.2).

The null result of Experiment 2 at least suggests that the greater effect of T2 than T1 on task performance in Experiment 1 is not caused by the +2 dB greater sound pressure level of T2 relative to T1 in that experiment. Similarly, the current results also suggest that that the decline in cognitive performance with increasing number of simultaneous talkers in multi-talker speech in Yadav et al. was not attributable to the accompanying increase in overall level [6]. In general, the results of this experiment are also consistent with previous findings, where it was the degree of changing-state of sound, and not the overall level, that affected the magnitude of the ISE in serial recall [22–25]. These previous findings, however, were based on relatively simplistic experimental designs with mostly fixed-location single-talker looped speech content, and limited semantic validity. Moreover, sound levels that are more representative of actual open-plan offices were used here compared to some previous studies where somewhat unrealistic sound levels (with respect to offices) were used [17,22]. The current findings extend the level-invariance in the ISE in serial recall to spatially-separated multi-talker environments with two simultaneously active talkers, at least for the 12 dB range tested within T2 sound environments. For the same range of levels and multi-talker considerations, the current findings introduce the finding of level-invariance for a writing task.

Despite the level-invariance of the ISE over a representative speech level range for open-plan offices shown here, the influence of any confounding factors that may affect the current null result needs to be considered further. Since the experimental design simply consisted of overall gain changes, it was not possible to determine whether dynamic changes to the SPL of some of the voices, or some finer changes to certain speech portions, may in fact, affect the ISE for both serial recall and writing tasks. However, the fairly comprehensive findings of Tremblay and Jones [24], where it was the changing-state, and not the overall level, or the level changes of individual elements (spoken digits) of the irrelevant speech that affected the ISE in serial recall, suggest that more detailed level changes may not change the current findings for serial recall; although the findings of Tremblay and Jones were based on simplistic sound environments [24]. In terms of semanticity, Westermann and Buchholz reported [40] informational masking for intelligible monologues from two spatially-separated, nearby talkers within an ecologically-valid multi-talker sound environment simulation of a cafeteria scenario. While their findings were based on a different experimental paradigm than the ISE, these findings do suggest that the effect of semantic cues may be somewhat reduced due to not just energetic, but also informational masking in multi-talker speech [40]. However, determining the extent to which semantic cues are curtailed within multi-talker speech, and whether fine-grained changes to SPL of speech, its spectrum, talker location, and their interaction, further affect the ISE in semantic tasks, or confound the current results, requires future research. In general, more work is suggested in the domain of complex tasks requiring semantic processing, since the performance in such tasks may not be immune to very large sound level changes (which may, however, be unrealistic for office scenarios) – potentially increasing instances of deviation of attention from focal processing.

### 3.3. General discussion

Although the current results can be generalized to similar settings where nearby talkers are intelligible and the ISE is relevant, it is worth considering the context of the experimental scenarios. The simulation here was specifically of hearing conditions in medium-sized open-plan office with a low background-noise of ~31 dBA (reliable levels for such spaces are not well-established in research literature), which allows focusing primarily on the speech-based effect. Moreover, similar background-noise levels were used in Haapakangas et al. [3] (~35 dBA), which facilitated comparisons with their results. Larger offices will typically have longer reverberation times than the current (mid-frequency $T_{30}$ = 0.3 s), and their interplay with background-noise is interesting in its own regard, where phenomena such as the Lombard effect [41] may lead to increased voice output by talkers. However, the speech from nearby talkers is still likely to be intelligible (Fig. 1) and perhaps level-invariant in the ISE (to an extent) [3,4,6], unless substantially masked due to an artificial masking system, reverberant unintelligible speech and/or general background-noise. Nevertheless, variations in broadband background-noise and room acoustics, along with the use of office sounds other than speech are recommended as factors for future studies, especially in relation to their potential as a speech masker.

The current results have some relevance for ISO 3382-3, which, despite being a predominantly room acoustic standard, does incorporate certain findings from psychoacoustics, and cognitive psychology literature – especially within the concept underlying distraction distance (see page 2, [2]), which was recently shown to be the most important predictor of speech issues in open-plan offices [1]. In its introduction, ISO 3382-3 ascribes maximally distracting status to single-talker scenarios, which appears to be an oversight, and should be amended in future versions of the standard. The single-talker assumption is not supported by current and previous results [4,6], and does not consider psychoacoustic benefits (or detriment in terms of increased ISE) of spatial separation of talkers in multi-talker environments [29]. In fact, even the source attributed in ISO 3382-3 (ref [10] therein) provided results that run counter to the single-talker assumption ([4]; abstract and Experiment 4 therein). Furthermore, a future version of ISO 3382-3 can highlight that multi-talker speech from nearby talkers may still distract, even after adequate acoustic treatment in accordance with the standard's methodology.

In general, the current results encourage studies that consider the complexities of multi-talker speech in open-plan offices, to provide more comprehensive tools to tackle speech distraction in these (i.e., going beyond ISO 3382-3), and similar settings. The ISE, as a system to study speech distraction, may also benefit from the more inclusive spectro-temporal-spatial considerations shown here, e.g. in characterizing babble speech for spatially-separated talkers [4].

### 4. Conclusions

1. Consistent with the ISE, the introduction of irrelevant speech in a simulated open-plan work environment reduces task performance.
2. Two simultaneous talkers are more distracting than one for serial recall task performance, which is consistent with previous studies [6,24], but also extends previous findings where single-talker speech was used [3].
3. A writing task is affected by ISE (with two simultaneous talkers): this is supported by some previous studies [5,12,13], whereas others found no effect [3]. The current study found no effect on writing performance for single talkers, but it did find an effect for two simultaneous talkers (new result).
4. ISE is level-invariant for two simultaneous talkers over a 12 dB range of overall speech levels that are representative of open-plan offices (new result).

### Acknowledgements

The authors thank Lucas Brooker, James Love and Jungsoo Kim for providing assistance with the experimental work in this study. This study was funded by the Australian Research Council's Discovery Projects funding scheme (project DP160103978). They also thank two anonymous reviewers for their comments, which helped to substantially improve the paper.

### References

[1] Haapakangas A, Hongisto V, Eerola M, Kuusisto T. Distraction distance and perceived disturbance by noise—an analysis of 21 open-plan offices. J Acoust Soc Am 2017;141:127–36.
[2] ISO. 3382-3 Acoustics – measurement of room acoustic parameters – Part 3: Open plan offices. Geneva, Switzerland: International Organization for Standardization; 2012.
[3] Haapakangas A, Hongisto V, Hyönä J, Kokko J, Keränen J. Effects of unattended speech on performance and subjective distraction: the role of acoustic design in open-plan offices. Appl Acoust 2014;86:1–16.
[4] Jones D, Macken W. Auditory babble and cognitive efficiency: role of number of voices and their location. J Exp Psychol Appl 1995.
[5] Banbury S, Berry DC. Disruption of office-related tasks by speech and office noise. Br J Psychol 1998;89:499–517.
[6] Yadav M, Kim J, Cabrera D, de Dear R. Auditory distraction in open-plan office environments: the effect of multi-talker acoustics. Appl Acoust 2017;126:68–80.
[7] Hughes RW. Auditory distraction: a duplex-mechanism account: Duplex-mechanism account of auditory distraction. PsyCh J 2014;3:30–41.
[8] Hughes RW, Marsh JE. The functional determinants of short-term memory: Evidence from perceptual-motor interference in verbal serial recall. J Exp Psychol Learn Mem Cogn 2017;43:537–51.
[9] Venetjoki N, Kaarlela-Tuomaala A, Keskinen E, Hongisto V. The effect of speech and speech intelligibility on task performance. Ergonomics 2006;49:1068–91.
[10] Keus van de Poll M, Ljung R, Odelius J, Sörqvist P. Disruption of writing by background speech: The role of speech transmission index. Appl Acoust 2014;81:15–8.
[11] Keus van de Poll M, Carlsson J, Marsh JE, Ljung R, Odelius J, Schlittmeier SJ, et al. Unmasking the effects of masking on performance: The potential of multiple-voice masking in the office environment. J Acoust Soc Am 2015;138:807–16.
[12] Sörqvist P, Nöstl A, Halin N. Disruption of writing processes by the semanticity of background speech: speech impairs writing processes. Scand J Psychol 2012;53:97–102.
[13] Ransdell S, Levy CM, Kellogg RT. The structure of writing processes as revealed by secondary task demands. L1-Educ Stud. Lang Lit 2002;2:141–63.

[14] Bell R, Röer JP, Marsh JE, Storch D, Buchner A. The effect of cognitive control on different types of auditory distraction: a preregistered study. Exp Psychol 2017;64:359–68.
[15] Marsh JE, Hughes RW, Jones DM. Auditory distraction in semantic memory: a process-based approach. J Mem Lang 2008;58:682–700.
[16] Bregman AS. Auditory scene analysis: the perceptual organization of sound. MIT press; 1994.
[17] Ellermeier W, Zimmer K. The psychoacoustics of the irrelevant sound effect. Acoust Sci Technol 2014;35:10–6.
[18] Buchner A, Irmen L, Erdfelder E. On the irrelevance of semantic information for the lIirrelevant speech effect. Q J Exp Psychol Sect A 1996;49:765–79.
[19] Buchner A, Rothermund K, Wentura D, Mehl B. Valence of distractor words increases the effects of irrelevant speech on serial recall. Mem Cognit 2004;32:722–31.
[20] Röer JP, Körner U, Buchner A, Bell R. Semantic priming by irrelevant speech. Psychon Bull Rev 2016.
[21] Marsh J, Jones D. Cross-modal distraction by background speech: what role for meaning? Noise Health Mumbai 2010;12:210–6.
[22] Ellermeier W, Hellbrück J. Is level irrelevant in "irrelevant speech"? Effects of loudness, signal-to-noise ratio, and binaural unmasking. J Exp Psychol Hum Percept Perform 1998;24:1406–14.
[23] Colle HA. Auditory encoding in visual short-term recall: effects of noise intensity and spatial location. J Verbal Learn Verbal Behav 1980;19:722–35.
[24] Tremblay S, Jones DM. Change of intensity fails to produce an irrelevant sound effect: implications for the representation of unattended sound. J Exp Psychol Hum Percept Perform 1999;25:1005–15.
[25] Schlittmeier SJ, Hellbrück J, Thaden R, Vorländer M. The impact of background speech varying in intelligibility: Effects on cognitive performance and perceived disturbance. Ergonomics 2008;51:719–36.
[26] NFS 31 199:2016 Acoustics - Acoustics Performances Of Open-Plan Offices 2016.
[27] Veitch JA, Bradley JS, Legault LM, Norcross S, Svec JM. Masking speech in open-plan offices with simulated ventilation noise: noise level and spectral composition effects on acoustic satisfaction. Inst Res Constr Intern Rep IRC-IR-846 2002.
[28] Zaglauer M, Drotleff H, Liebl A. Background babble in open-plan offices: a natural masker of disruptive speech? Appl Acoust 2017;118:1–7.
[29] Bronkhorst AW. The cocktail-party problem revisited: early processing and selection of multi-talker speech. Atten Percept Psychophys 2015;77:1465–87.
[30] Reinten J, Braat-Eggen PE, Hornikx M, Kort HSM, Kohlrausch A. The indoor sound environment and human task performance: A literature review on the role of room acoustics. Build Environ 2017;123:315–32.
[31] Rindel JH. The use of computer modeling in room acoustics. J Vibroengineering 2000;3..
[32] Vigeant MC, Wang LM, Rindel JH. Objective and subjective evaluations of the multi-channel auralization technique as applied to solo instruments. Appl Acoust 2011;72:311–23.
[33] ANSI 3.5-1997 (R2012). American National Standard – Methods for Calculation of the Speech Intelligibility Index.
[34] Algazi VR, Duda RO, Thompson DM, Avendano C. The cipic hrtf database. Appl. Signal Process. Audio Acoust. 2001 IEEE Workshop On, IEEE 2001;2001:99–102.
[35] Schlittmeier SJ, Liebl A. The effects of intelligible irrelevant background speech in offices – cognitive disturbance, annoyance, and solutions. Facilities 2015;33:61–75.
[36] Jørgensen S, Ewert SD, Dau T. A multi-resolution envelope-power based model for speech intelligibility. J Acoust Soc Am 2013;134:436.
[37] Schutte M, Marks A, Wenning E, Griefahn B. The development of the noise sensitivity questionnaire. Noise Health 2007;9:15–24.
[38] Pinheiro J, Bates D, DebRoy S, Sarkar D. R Core Team (2014) nlme: linear and nonlinear mixed effects models. R package version 3.1-117.
[39] Lenth RV, Hervé M. lsmeans: Least-Squares Means. R package version 2.27-61; 2018.
[40] Westermann A, Buchholz JM. The influence of informational masking in reverberant, multi-talker environments. J Acoust Soc Am 2015;138:584–93.
[41] Nijs L, Saher K, den Ouden D. Effect of room absorption on human vocal output in multitalker situations. J Acoust Soc Am 2008;123:803.